\documentclass[twocolumn,amsmath,amssymb,aps,showpacs]{revtex4-2}
\usepackage[utf8]{inputenc}
\usepackage[T1]{fontenc}
\usepackage{graphicx}
\usepackage{dcolumn}
\usepackage{bm}
\usepackage{hyperref}
\usepackage{xcolor}
\usepackage{physics}
\usepackage{microtype}
\usepackage{booktabs}
\usepackage{array}
\usepackage{amssymb}
\usepackage{newtxtext,newtxmath}

\hypersetup{
    colorlinks=true,
    linkcolor=blue,
    filecolor=magenta,
    urlcolor=cyan,
    citecolor=blue,
}

\begin{document}

\title{The Soul of Waves: Physical Interpretation of Dispersion Relations}

\author{Renato Vieira dos Santos}
\email{renato.santos@ufla.br}
\affiliation{Instituto de Ci\^{e}ncia, Tecnologia e Inova\c{c}\~{a}o - ICTIN, Universidade Federal de Lavras - UFLA, Campus Para\'{i}so, MG 37950-000, Brazil}

\date{\today}

\begin{abstract}
This pedagogical paper presents a comprehensive framework for interpreting dispersion relations across fundamental physical systems. We adopt a novel approach that starts from the mathematical form $\omega(\mathbf{k})$ and systematically extracts its physical content, rather than deriving it from first principles. Through an in-depth case study of the massive Klein-Gordon dispersion relation $\omega^2 = \omega_0^2 + c^2k^2$, we demonstrate how this single equation encodes phase velocity, group velocity, density of states, effective mass, and impedance. The analysis reveals the universal nature of this dispersion form, which manifests in quantum fields, plasmas, superconductors, and photonic crystals with different physical interpretations of its parameters. We complement this with detailed examination of classical systems including mass-spring chains and hydrodynamic waves, providing tangible analogies that bridge conceptual understanding between quantum and classical wave phenomena. The paper includes eleven carefully designed figures that visualize key concepts and a comprehensive catalog of dispersion relations in the Appendix. Aimed at advanced undergraduates and instructors, this work emphasizes conceptual understanding through physical interpretation, offering a unified pedagogical framework for teaching wave propagation across physics curricula while maintaining mathematical clarity and depth.

\vspace{0.5em}


\vspace{0.5em}

Este artigo pedagógico apresenta uma estrutura abrangente para interpretar relações de dispersão em sistemas físicos fundamentais. Adotamos uma abordagem inovadora que parte da forma matemática $\omega(\mathbf{k})$ e da qual se extrai sistematicamente seu conteúdo físico, em vez de derivá-la a partir de primeiros princípios. Através de um estudo de caso detalhado da relação de dispersão de Klein-Gordon massivo $\omega^2 = \omega_0^2 + c^2k^2$, demonstramos como esta única equação codifica velocidade de fase, velocidade de grupo, densidade de estados, massa efetiva e impedância. A análise revela a natureza universal desta forma de dispersão, que se manifesta em campos quânticos, plasmas, supercondutores e cristais fotônicos com diferentes interpretações físicas de seus parâmetros. Complementamos isto com um exame detalhado de sistemas clássicos, incluindo cadeias massa-mola e ondas hidrodinâmicas, fornecendo analogias tangíveis que conectam a compreensão conceitual entre fenômenos ondulatórios quânticos e clássicos. O artigo inclui onze figuras cuidadosamente elaboradas que visualizam conceitos-chave e um catálogo abrangente de relações de dispersão no Apêndice. Voltado para alunos de graduação avançados e instrutores, este trabalho enfatiza a compreensão conceitual através da interpretação física, oferecendo uma estrutura pedagógica unificada para ensinar propagação de ondas em currículos de física, mantendo clareza e profundidade matemáticas.

\vspace{0.5em}

\noindent\textbf{Palavras-chave:} relações de dispersão, propagação de ondas, massa efetiva, densidade de estados, velocidade de fase, velocidade de grupo.

\end{abstract}

\maketitle

\section{Introduction: The Universal Language of Dispersion Relations}

Imagine a physicist's toolbox containing a single instrument capable of revealing how light bends through a prism, how electrons dance through crystals, how tsunamis cross oceans, and how the Higgs boson acquires mass. This improbable multi-tool exists: it is the dispersion relation $\omega(\mathbf{k})$, the functional connection between a wave's temporal rhythm ($\omega$) and its spatial pattern ($\mathbf{k}$). In this mathematical relationship resides a remarkable synthesis, for as different as these phenomena appear, they all speak the common language of wave propagation \cite{Whitham1974}.

Consider a stone dropped in a pond. The large, slow swells and the tiny, rapid ripples that emanate from the impact obey different rules: gravity governs the former, surface tension the latter. Yet both are described by a single equation $\omega^2 = gk + (\sigma/\rho)k^3$, where the balance between the linear and cubic terms encodes which force dominates at each scale. This is the essence of a dispersion relation: it tells us not only how waves travel but why they travel that way, revealing the underlying physical forces at play.

Our journey begins with perhaps the most elegant example, the Klein-Gordon dispersion relation:
\begin{equation}
\omega^2 = \omega_0^2 + c^2k^2,
\label{eq:KG_dispersion_intro}
\end{equation}
where $\omega_0 = mc^2/\hbar$ for a relativistic particle of mass $m$. What does this equation describe physically? The story starts with Einstein's relativistic energy-momentum relation $E^2 = p^2c^2 + m^2c^4$. In quantum mechanics, energy and momentum become operators: $E \rightarrow i\hbar\partial_t$, $\mathbf{p} \rightarrow -i\hbar\nabla$. Substituting these into the energy-momentum relation yields the Klein-Gordon equation for a scalar field $\psi(\mathbf{r},t)$:
\begin{equation}
\left(-\frac{\hbar^2}{c^2}\frac{\partial^2}{\partial t^2} + \hbar^2\nabla^2 - m^2c^2\right)\psi(\mathbf{r},t) = 0.
\label{eq:KG_operator}
\end{equation}
The field $\psi(\mathbf{r},t)$ might represent pions mediating nuclear forces or the Higgs field endowing particles with mass \cite{Bjorken1965,Itzykson1980}. Seeking wave-like solutions $\psi(\mathbf{r},t) = \psi_0 e^{i(\mathbf{k}\cdot\mathbf{r}-\omega t)}$ transforms this partial differential equation into an algebraic condition: $(-\hbar^2\omega^2/c^2 + \hbar^2k^2 + m^2c^2)\psi_0 = 0$. For non-vanishing $\psi_0$, we recover exactly Eq.~\eqref{eq:KG_dispersion_intro}. Thus, the dispersion relation emerges, revealing which wave patterns ($\omega$, $\mathbf{k}$) the physical system permits.

Remarkably, this same mathematical form $\omega^2 = \omega_0^2 + c^2k^2$ appears in contexts far removed from quantum field theory. In a plasma, $\omega_0$ becomes the plasma frequency $\omega_p$, below which electromagnetic waves cannot propagate \cite{Chen2016}. In a superconductor, $\omega_0 = c/\lambda_L$ where $\lambda_L$ is the London penetration depth, 
explaining why magnetic fields are expelled (the Meissner effect) \cite{London1935,Tinkham2004}. In photonic crystals, $\omega_0$ marks a band edge where light waves transition from propagating to evanescent \cite{Yablonovitch1987,Joannopoulos2011}. The mathematics remains identical while the physics changes costume.

This universality makes dispersion relations powerful pedagogical tools. From $\omega(\mathbf{k})$ alone, we can extract:
\begin{enumerate}
\item Phase velocity $v_p = \omega/k$: how wave crests appear to move
\item Group velocity $v_g = d\omega/dk$: how energy and information actually travel
\item Density of states $g(\omega)$: how many distinct wave modes exist at each frequency
\item Effective mass $m_{\text{eff}} = \hbar/(d^2\omega/dk^2)$: how waves respond to forces
\item Impedance $Z(\omega)$: how waves reflect and transmit at boundaries
\end{enumerate}

Each of these quantities emerges from mathematical features of the $\omega(\mathbf{k})$ curve: its slope, curvature, intercepts, and asymptotes \cite{DifracaoDispersao}. The dispersion relation thus serves as a Rosetta Stone, translating between the abstract mathematics of wave equations and tangible physical behavior \cite{Jackson1998}.

In this pedagogical paper, we reverse the traditional approach. Rather than deriving dispersion relations from physical principles, we begin with $\omega(\mathbf{k})$ as given and systematically excavate its physical content. We conduct a deep case study of the Klein-Gordon form \eqref{eq:KG_dispersion_intro}, demonstrating how every mathematical feature corresponds to measurable physics. We then extend this interpretative framework to classical systems: mass-spring chains that illustrate Brillouin zones and band gaps, and water waves that reveal competition between gravity and surface tension. Through eleven carefully designed figures and a comprehensive catalog of dispersion relations in the Appendix \ref{sec:appendix}, we provide both conceptual understanding and practical tools for interpreting wave phenomena across physics.

Our goal is to equip students with what might be called ``dispersion relation literacy'', the ability to look at a plot of $\omega$ versus $k$ and read from it the story of how waves behave in that system. Whether in quantum mechanics, solid state physics, optics, or fluid dynamics, this literacy transforms dispersion relations from calculation tools into cognitive frameworks for understanding wave propagation in its many guises.

\section{The Klein-Gordon Dispersion: A Universe in an Equation}

What do quantum fields, plasmas, superconductors, and photonic crystals fundamentally share? At first consideration, these systems appear profoundly different. Yet they all obey the same mathematical relation when waves travel through them: $\omega^2 = \omega_0^2 + c^2k^2$. This recurrence of identical mathematical structure across vastly different physical domains is a striking example of the unifying power of dispersion relations.

Before delving into the rich interpretations encoded in dispersion relations, we clarify what a dispersion relation fundamentally represents. In wave physics, we typically seek solutions of the form $\psi(\mathbf{r}, t) = A e^{i(\mathbf{k}\cdot\mathbf{r} - \omega t)}$, where $\mathbf{k}$ is the wave vector quantifying spatial frequency and $\omega$ is the angular frequency quantifying temporal frequency. The dispersion relation constitutes the equation connecting these quantities, emerging inevitably from the physical laws governing wave propagation in the specific system. 

Now, here's a subtle point that often trips up students: should we write $\omega(k)$ or $k(\omega)$? As a pedagogical analogy (not a fundamental symmetry), one might think of it like asking whether distance depends on time or time depends on distance when driving a car. Mathematically, either function works, but physically, one perspective often makes more intuitive sense. The $\omega(k)$ view asks: ``If I freeze a wave's spatial pattern (fix $k$), what oscillation frequency will it naturally have?'' This perspective suits quantum mechanics, where particles have definite momentum ($p = \hbar k$) and we want their corresponding energy. The $k(\omega)$ view asks: ``If I shake something at frequency $\omega$, what spatial pattern will emerge?'' This fits wave experiments where we control the driving frequency and observe the response \cite{Whitham1974}. (We emphasize that this is a pedagogical analogy; unlike the relation between position and time in Newtonian physics, $\omega$ and $k$ are on equal footing as Fourier variables.)

\begin{figure}[htbp]
\centering
\includegraphics[width=\columnwidth]{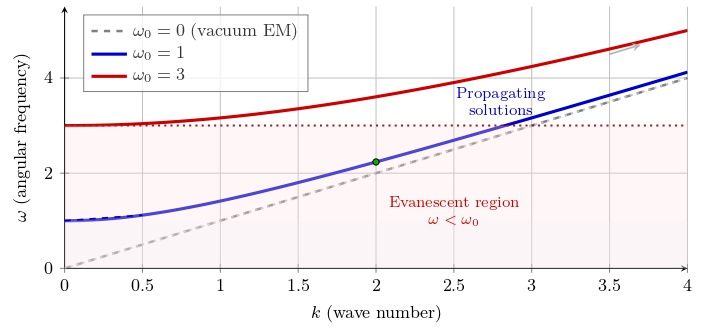}
\caption{The massive Klein-Gordon dispersion relation $\omega^2 = \omega_0^2 + c^2 k^2$ illustrated for three representative cases: $\omega_0 = 0$ (black dashed, vacuum electromagnetic limit), $\omega_0 = 1\times10^{15}\,\text{rad/s}$ (blue, typical optical frequency gap), and $\omega_0 = 3\times10^{15}\,\text{rad/s}$ (red, larger gap). The gray dotted line shows $\omega = ck$ for comparison. A finite $\omega_0$ imposes a minimum frequency for wave propagation, with imaginary $k$ solutions for $\omega < \omega_0$ indicating evanescent decay rather than propagation. This universal mathematical form manifests in diverse physical contexts: plasmas ($\omega_0 = \omega_p$) \cite{Chen2016}, superconductors ($\omega_0 = c/\lambda_L$) \cite{London1935,Tinkham2004}, photonic crystals near band edges \cite{Yablonovitch1987,Joannopoulos2011}, and relativistic quantum mechanics ($\omega_0 = mc^2/\hbar$) \cite{Bjorken1965}. The curves illustrate how increasing $\omega_0$ raises the minimum frequency while preserving the asymptotic linear behavior at large $k$.}
\label{fig:klein_gordon_dispersion}
\end{figure}

Consider then the dispersion relation emerging from the Klein-Gordon equation, already introduced in Eq.~\eqref{eq:KG_dispersion_intro}. At first glance, this equation seems almost too simple to be interesting---a sum of two squares. Yet this deceptively simple form encodes an extraordinary range of physical behaviors. Figure~\ref{fig:klein_gordon_dispersion} reveals the rich structure hidden within this formula.

The most distinctive feature—the $\omega_0^2$ term—acts like a ``frequency floor.'' Imagine trying to make a very stiff trampoline bounce: there's a minimum force you must apply before it responds at all. Below that threshold, you just get a dull thud; above it, proper bouncing begins. This $\omega_0$ parameter wears different costumes in different systems: in quantum field theory, it's the particle's mass ($\omega_0 = mc^2/\hbar$) \cite{Peskin1995}; in plasmas, it's the collective electron dance frequency ($\omega_0 = \omega_p$) \cite{Chen2016}; in superconductors, it's related to how deeply magnetic fields can penetrate ($\omega_0 = c/\lambda_L$) \cite{Tinkham2004}. Same mathematics, different physical actors.

From the $\omega(k)$ perspective natural to quantum systems, $\omega_0$ represents the energy at rest: when $k = 0$ (particle at rest), $\omega = \omega_0$ gives $E = \hbar\omega_0 = mc^2$. 
This is like measuring a car's minimum fuel consumption when it's idling. Even doing nothing, it needs some energy just to exist. The vacuum, by contrast ($\omega = ck$), has zero idle cost: $k=0$ means $\omega=0$, complete stillness. The $\omega_0$ term thus embodies an intrinsic ``vibration'' that persists even without spatial variation.

Now shift to the $k(\omega)$ perspective appropriate for wave propagation experiments. Here the physics becomes particularly vivid: when you try to send a wave with frequency $\omega$ through the medium, the medium responds with wave number $k(\omega) = \pm\frac{1}{c}\sqrt{\omega^2 - \omega_0^2}$. If $\omega < \omega_0$, $k$ becomes imaginary. In ordinary life, we rarely encounter imaginary distances, but in wave physics, imaginary $k$ means exponential decay: $k = i\kappa$ gives solutions like $e^{-\kappa x}$ instead of $e^{ikx}$.

Think of it this way: sending a low-frequency wave ($\omega < \omega_0$) into such a medium is like trying to whistle in a room with perfect acoustic absorption. You blow air (energy in), but instead of a clear whistle (propagating wave), you get a muffled sigh that dies immediately near your lips (exponential decay). This explains why radio waves below the plasma frequency bounce off the ionosphere \cite{Chen2016}, and why magnetic fields can't penetrate far into superconductors \cite{Tinkham2004}. The materials are ``acoustically absorbent'' at those frequencies.

\begin{figure}[htbp]
\centering
\includegraphics[width=\columnwidth]{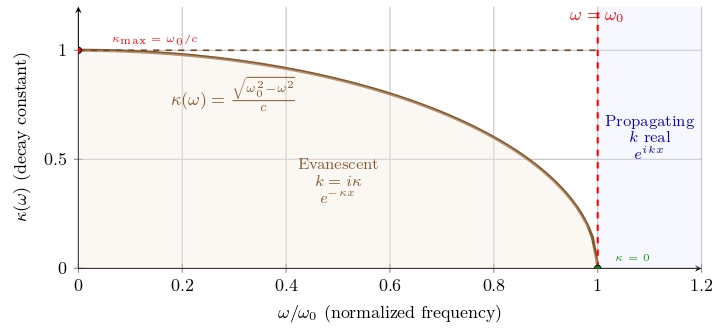}
\caption{Decay constant $\kappa(\omega) = \sqrt{\omega_0^2 - \omega^2}/c$ for frequencies below the gap ($\omega < \omega_0$) with $\omega_0 = 1\times10^{15}\,\text{rad/s}$. This figure adopts the $k(\omega)$ perspective relevant for wave propagation experiments: given a driving frequency $\omega$, what decay constant $\kappa = \text{Im}[k]$ characterizes the medium's response? When $\omega < \omega_0$, the wave number becomes imaginary $k = i\kappa$, yielding exponentially decaying solutions $e^{-\kappa x}$ rather than propagating waves $e^{ikx}$. The decay constant varies from $\kappa_{\text{max}} = \omega_0/c$ at $\omega = 0$ (strongest decay) to $\kappa = 0$ at $\omega = \omega_0$ (critical point). This evanescent behavior explains multiple physical phenomena: magnetic field penetration in superconductors (Meissner effect, $\kappa^{-1} = \lambda_L$) \cite{Tinkham2004}, radio wave reflection from the ionosphere \cite{Chen2016}, and frustrated total internal reflection in optics \cite{Joannopoulos2011}. The blue shaded region ($\omega > \omega_0$) indicates propagating solutions with real $k$. The smooth variation of $\kappa$ with $\omega$ demonstrates how the system transitions continuously from strong localization to propagation as frequency increases through $\omega_0$.}
\label{fig:evanescent_decay}
\end{figure}

Figure~\ref{fig:evanescent_decay} explicitly adopts this $k(\omega)$ perspective, plotting $\kappa(\omega) = \text{Im}[k(\omega)]$ against normalized frequency. Notice how $\kappa$ starts high at $\omega = 0$ (strong absorption, quick decay) and gradually decreases to zero at $\omega = \omega_0$ (threshold of propagation). This smooth variation illustrates the continuous transition from strong spatial localization to full propagation as the driving frequency increases through $\omega_0$.

The complementary $\omega(k)$ perspective, shown in Fig.~\ref{fig:klein_gordon_dispersion}, reveals the system's natural spectrum. Each point on the curve represents an allowed plane wave mode with specific $k$ and corresponding $\omega$. The gap appears here as a minimum frequency $\omega_0$ that all modes must exceed---frequencies below this threshold simply do not exist as propagating solutions.

Thus, the Klein-Gordon relation teaches us to choose our perspective like choosing the right tool: $\omega(k)$ for understanding the natural oscillation modes of a system, $k(\omega)$ for predicting its response to external driving. Both perspectives are valid and complementary. The mathematics $\omega^2 = \omega_0^2 + c^2k^2$ contains both stories simultaneously, an elegant economy that explains why this single equation appears everywhere from the cosmic (Higgs field) to the everyday (why your radio sometimes loses signal).

\subsection{Phase and Group Velocities: Two Fundamental Speeds}

If waves were participants in a race, they would possess two distinct speeds: one governing their visual appearance and another governing their physical substance. The crests of a wave can appear to move faster than light in vacuum, while the actual information and energy propagate at more modest speeds. This fundamental duality emerges naturally from our Klein-Gordon equation.

\begin{figure}[htbp]
\centering
\includegraphics[width=\columnwidth]{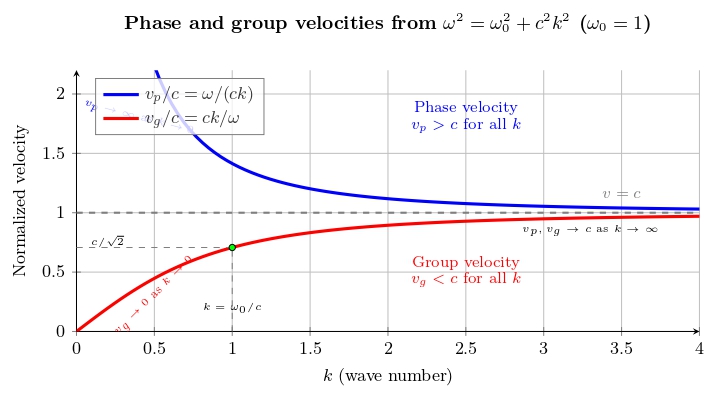}
\caption{Normalized phase velocity $v_p/c = \omega/(ck)$ (blue curve) and group velocity $v_g/c = ck/\omega$ (red curve) derived from the dispersion relation $\omega^2 = \omega_0^2 + c^2 k^2$ with $\omega_0 = 1\times10^{15}\,\text{rad/s}$. The phase velocity exceeds $c$ for all finite $k$, approaching infinity as $k \to 0$ and asymptotically approaching $c$ as $k \to \infty$, but carries no energy or information. The group velocity, representing signal propagation speed, remains subluminal ($v_g < c$) for all $k$, vanishing as $k \to 0$ and approaching $c$ as $k \to \infty$. At the specific wave number $k = \omega_0/c$, the group velocity equals $v_g = c/\sqrt{2} \approx 0.707c$. The inequality $v_p > c > v_g$ preserves relativistic causality while allowing wave crests to move faster than light. The curves illustrate how phase and group velocities approach equality only in the high-frequency limit where $\omega \gg \omega_0$.}
\label{fig:phase_group_velocities}
\end{figure}

The phase velocity $v_p = \omega/k$ describes how rapidly wave crests appear to travel. For the Klein-Gordon dispersion (Eq.~\eqref{eq:KG_dispersion_intro}),
\begin{equation}
v_p = \frac{\omega}{k} = c\sqrt{1 + \frac{\omega_0^2}{c^2k^2}},
\end{equation}
invariably exceeds $c$, as shown in Fig.~\ref{fig:phase_group_velocities}. This superluminal appearance doesn't violate relativistic causality because wave crests, like mirages, can move arbitrarily fast without transporting energy or information \cite{Brillouin1960}.

The physically significant propagation speed is the group velocity
\begin{equation}
v_g = \frac{d\omega}{dk} = \frac{c^2k}{\omega} = c\sqrt{1 - \frac{\omega_0^2}{\omega^2}},
\label{eq:group_velocity_derived}
\end{equation}
which remains subluminal. This quantity describes how energy, information, and wave packets move through the medium. When $\omega$ approaches $\omega_0$ from above, $v_g$ approaches zero, creating ``slow light'' phenomena in photonic crystals or sluggish radio wave propagation near the plasma frequency \cite{Baba2008}.

A pulse with finite bandwidth $\Delta\omega$ contains frequency components that travel at different group velocities. This variation of $v_g$ with frequency---not the mere fact that $v_f \neq v_g$---is what causes wave packets to spread during propagation. If the group velocity were constant (i.e., $d^2\omega/dk^2 = 0$), all spectral components would travel together and the pulse would maintain its shape indefinitely. The spreading rate is quantified by the group velocity dispersion
\begin{equation}
\frac{d^2\omega}{dk^2} = \frac{c^2}{\omega}\left(1 - \frac{c^2k^2}{\omega^2}\right) = \frac{c^2\omega_0^2}{\omega^3},
\label{eq:group_velocity_dispersion}
\end{equation}
always positive for $\omega > \omega_0$, indicating normal dispersion where higher-frequency components travel faster than lower-frequency ones. The pulse thus broadens over time, analogous to runners in a marathon gradually spreading out due to their differing speeds. This connection between dispersion curvature and pulse broadening is one of the five key physical quantities encoded in the shape of $\omega(k)$, alongside the phase and group velocities themselves.

\begin{figure}[htbp]
\centering
\includegraphics[width=\columnwidth]{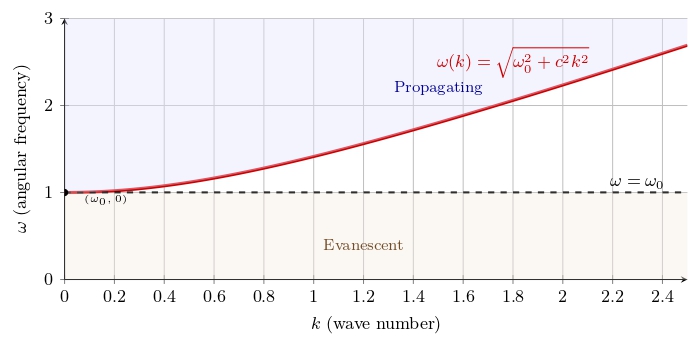}
\caption{Phase diagram in the $(\omega, k)$ plane for the dispersion relation $\omega^2 = \omega_0^2 + c^2k^2$ with $\omega_0 = 1\times10^{15}\,\text{rad/s}$. The red curve shows the dispersion relation $\omega(k) = \sqrt{\omega_0^2 + c^2k^2}$. Blue region: propagating solutions with real $k$ and oscillatory spatial behavior $e^{ikx}$. Brown region: evanescent solutions with imaginary $k = i\kappa$ and exponential spatial decay $e^{-\kappa x}$ for $\omega < \omega_0$. The horizontal dashed line at $\omega = \omega_0$ marks the frequency gap boundary. No physically allowed propagating solutions exist in the white region between the brown evanescent region and the dispersion curve (where $\omega_0 < \omega < \sqrt{\omega_0^2 + c^2k^2}$). This diagram provides a complete map of wave behavior: given any $(\omega, k)$ pair, one can immediately determine whether the solution propagates or decays exponentially. The diagram visualizes how the system transitions smoothly from evanescent to propagating behavior as frequency increases through $\omega_0$.}
\label{fig:phase_diagram}
\end{figure}

\subsection{Density of States: Counting the Available Wave Modes}

The density of states $g(\omega)$ answers a deceptively simple question: \textit{at a given frequency $\omega$, how many distinct plane wave solutions $e^{i(\mathbf{k}\cdot\mathbf{r}-\omega t)}$ does the system support?}

Conceptually, $g(\omega)$ functions like a \textit{linear density} along a one-dimensional axis. Imagine a long, straight highway with mile markers. At each mile, we ask: how many gas stations are there per mile? That is a linear density. For waves, the frequency axis $\omega$ is our highway, and $g(\omega)$ tells us how densely packed the available oscillation modes are at each point along this axis.

A low density means few modes per unit frequency — the system has limited ``options'' to oscillate at those frequencies. A high density means many modes packed closely together — a rich variety of vibrational possibilities. When $g(\omega) = 0$, the highway has no gas stations for that entire stretch: no modes exist at those frequencies, and waves simply cannot propagate.

\paragraph{What does $g(\omega)$ reveal for the Klein-Gordon relation?}

\begin{figure}[htbp]
\centering
\includegraphics[width=\columnwidth]{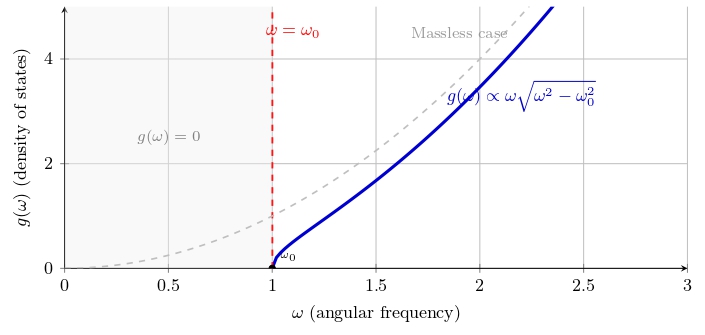}
\caption{Density of states $g(\omega)$ derived from the Klein-Gordon dispersion relation $\omega^2 = \omega_0^2 + c^2k^2$. For $\omega > \omega_0$, we obtain $g(\omega) = \frac{V}{2\pi^2 c^3} \omega\sqrt{\omega^2 - \omega_0^2}$ (blue curve), which vanishes at the gap edge $\omega = \omega_0$ and increases monotonically thereafter. The gray dashed curve shows the massless case $g(\omega) \propto \omega^2$, which the massive result approaches asymptotically for $\omega \gg \omega_0$. The red dashed line at $\omega = \omega_0$ marks the threshold frequency below which $g(\omega) = 0$, corresponding to the complete absence of propagating states. This characteristic density of states behavior explains multiple physical phenomena: suppression of spontaneous emission in photonic crystals near band edges, temperature dependence of heat capacity in gapped quantum systems, and spectral properties of massive quantum fields. The square-root onset at $\omega_0$ represents a universal feature of three-dimensional systems near band edges.}
\label{fig:density_of_states}
\end{figure}

For the dispersion relation $\omega^2 = \omega_0^2 + c^2k^2$ in three dimensions, the density of states takes a characteristic form (see Appendix~\ref{app:dos_derivation} for the detailed derivation):
\begin{equation}
g(\omega) = \frac{V}{2\pi^2 c^3} \, \omega\sqrt{\omega^2 - \omega_0^2}, \quad \text{for } \omega > \omega_0,
\label{eq:density_of_states_final}
\end{equation}
and $g(\omega) = 0$ for $\omega < \omega_0$, as shown in Fig.~\ref{fig:density_of_states}. Let us interpret this result physically, paying close attention to what it says about the \textit{density} of states.

\paragraph{Below $\omega_0$: zero density — a forbidden region.}

If $\omega < \omega_0$, then $g(\omega) = 0$. This means: \textbf{there are no propagating wave modes whatsoever in this frequency range}. It is like searching for trees in a desert — the density is zero. The system simply cannot oscillate at these frequencies as traveling waves.

This is the meaning of a \textit{band gap}: a region of the spectrum where the density of states vanishes. Nothing propagates. The wave cannot ``exist'' as a propagating wave.

\paragraph{Just above $\omega_0$: low density — the forest edge.}

For $\omega$ slightly greater than $\omega_0$, expanding Eq.~\eqref{eq:density_of_states_final} gives:
\begin{equation}
g(\omega) \approx \frac{V}{2\pi^2 c^3} \, \omega_0 \sqrt{2\omega_0(\omega - \omega_0)}.
\end{equation}
The density rises from zero like the square root of the distance from threshold. It is like a forest that begins with sparse trees at its edge and becomes denser as one moves inward.

This smooth onset has profound consequences. If you try to excite the system at a frequency very close to $\omega_0$, the system ``sees'' very few available modes. In a photonic crystal, for example, an excited atom in this frequency range has great difficulty emitting light --- spontaneous emission is strongly suppressed \cite{Yablonovitch1987,John1987}.

\paragraph{High frequencies: high density — dense forest.}

For $\omega \gg \omega_0$, the expression approaches the massless case:
\begin{equation}
g(\omega) \approx \frac{V}{2\pi^2 c^3} \, \omega^2.
\end{equation}
The density grows with the square of frequency: the higher the frequency, the more modes fit into the same interval. In this limit, the effect of the mass (the $\omega_0$ term) becomes irrelevant, and the system behaves like ordinary waves in vacuum.

\paragraph{Why density matters more than total count.}

The density $g(\omega)$ is a \textit{local} function: it tells us what happens \textit{at that specific frequency}. The total count $N(\omega)$ (the number of modes up to $\omega$) is cumulative information that hides fine details. The density reveals the spectral structure: where there are many modes (active regions), where there are few (transition regions), and where there are none (gaps).

In condensed matter physics, quantum optics, and quantum field theory, knowing the density of states is as essential as knowing, on a map, where there are mountains, where there are valleys, and where there are plains. It is the topography of the wave world. From this simple expression $g(\omega) \propto \omega\sqrt{\omega^2 - \omega_0^2}$, we can understand the suppression of spontaneous emission in photonic crystals \cite{Yablonovitch1987,John1987}, the temperature dependence of heat capacity in gapped quantum systems \cite{Debye1912,Reif1965}, and the spectral properties of massive quantum fields \cite{Weinberg1995}.

\subsection{Effective Mass: When Waves Acquire Inertia}

Waves governed by the Klein-Gordon dispersion acquire an effective mass that makes them respond to perturbations as if possessing material inertia. Expanding Eq.~\eqref{eq:KG_dispersion_intro} for small $k$ (recalling the non-relativistic kinetic energy $\hbar\omega \approx \hbar\omega_0 + \hbar^2 k^2/2m$):
\begin{equation}
\omega = \sqrt{\omega_0^2 + c^2k^2} \approx \omega_0 + \frac{c^2k^2}{2\omega_0} = \omega_0 + \frac{\hbar k^2}{2m_{\text{eff}}},
\end{equation}
yields the effective mass $m_{\text{eff}} = \hbar\omega_0/c^2$. This $m_{\text{eff}}$ plays the same dynamical role as inertial mass in Newton's second law, determining how waves respond to forces, accelerate in potential gradients, and interact with barriers.

The implications span physics subdisciplines. In quantum field theory, $m_{\text{eff}}$ returns the particle's rest mass: $m = \hbar\omega_0/c^2$. In superconductors, photons acquire mass $m_{\text{eff}} = \hbar/(c\lambda_L)$ through the Anderson-Higgs mechanism, explaining the Meissner effect \cite{London1935,Tinkham2004}: magnetic fields are expelled because photons become too ``heavy'' to penetrate deeply \cite{London1935}. In photonic crystals near band edges, light can exhibit negative effective mass, accelerating opposite to applied forces \cite{Joannopoulos2011}. In plasmas, electromagnetic waves below $\omega_p$ can be interpreted as having an infinite effective mass, signaling an inability to propagate as a traveling wave, analogous to a particle with infinite inertia.

\begin{figure}[htbp]
\centering
\includegraphics[width=\columnwidth]{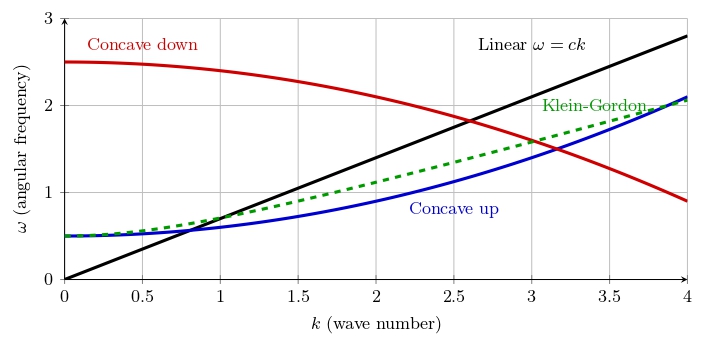}
\caption{Comparison of dispersion relations with different curvatures, illustrating fundamental wave propagation behaviors. Linear dispersion $\omega = ck$ (black) yields constant and equal phase and group velocities ($v_p = v_g = c$), representing non-dispersive propagation. Upward curvature $\omega \approx \omega_0 + \alpha k^2$ (blue, concave up) yields group velocity exceeding phase velocity ($v_g > v_p$), characteristic of anomalous dispersion regions. Downward curvature $\omega \approx \omega_0 - \beta k^2$ (red, concave down) gives $v_g < v_p$, typical near band edges in periodic systems. The Klein-Gordon form $\omega = \sqrt{\omega_0^2 + c^2k^2}$ (green dashed) shows monotonically decreasing slope, representing normal dispersion with $v_g$ always subluminal. Curvature sign determines wave packet spreading characteristics and whether higher frequencies travel faster (normal dispersion) or slower (anomalous dispersion) than lower frequencies. The second derivative $d^2\omega/dk^2$ provides the effective mass through $m_{\text{eff}} = \hbar/(d^2\omega/dk^2)$.}
\label{fig:curvature_comparison}
\end{figure}

Fig.~\ref{fig:curvature_comparison} illustrates how different dispersion curvatures correspond to different effective mass behaviors. The phase diagram in Fig.~\ref{fig:phase_diagram} reveals richer dynamics through this lens: near $\omega \approx \omega_0$, the flat dispersion ($d\omega/dk \approx 0$) corresponds to $m_{\text{eff}} \to \infty$, creating localized states and ``slow light''; far from the gap, near-linearity yields $m_{\text{eff}} \to 0$, producing weightless propagation at speed $c$.

The general relationship
\begin{equation}
\frac{1}{m_{\text{eff}}} = \frac{1}{\hbar} \frac{d^2\omega}{dk^2}
\end{equation}
applies universally across semiconductors, photonic crystals, superlattices, and quantum fluids. All can be described by Newtonian dynamics with an effective mass extracted from dispersion curvature. Thus, when encountering any dispersion relation, students should examine its second derivative: large positive $d^2\omega/dk^2$ indicates light quasiparticles; $d^2\omega/dk^2 \approx 0$ implies infinite mass and localization; negative $d^2\omega/dk^2$ signifies particles accelerating opposite to applied forces. From the Meissner effect to slow light to semiconductor electronics, this physics emerges encoded within the simple geometry of $\omega(k)$ curves.

\subsection{Generalized Wave Impedance: An Analogical Bridge}
\label{sec:impedance_analogy}

The term \textit{impedance} carries precise, system-specific definitions in different branches of physics: the electrical impedance \(Z=V/I\) in circuit theory, the intrinsic impedance \(\eta_0=\sqrt{\mu_0/\epsilon_0}\) of free space in electromagnetism, and the mechanical impedance \(F/v\) in acoustics. Here we adopt a broader mathematical perspective — the \textit{wave impedance} — defined as the ratio of a force-like field variable to a velocity-like field variable in a freely propagating wave. This pedagogical device is not an attempt to redefine the impedance of quantum fields, but rather to highlight a structural analogy that unifies the treatment of reflection and transmission across diverse wave systems \cite{Jackson1998,Pozar2011}. For the Klein–Gordon scalar field, the resulting expression should be regarded as an analogical extension, not a standard physical observable. Its value lies in revealing the common mathematical architecture shared by classical and quantum wave equations.

\paragraph{Warm-up: Impedance of a Stretched String.}

Consider a string under tension $T$ with linear mass density $\mu$. The wave equation is:
\[
\frac{\partial^2 y}{\partial t^2} - \frac{T}{\mu}\frac{\partial^2 y}{\partial x^2} = 0.
\]
For a harmonic wave $y(x,t) = y_0 e^{i(kx - \omega t)}$, the transverse force exerted by the left segment on the right segment at any point is $F = -T\,\partial y/\partial x$. The transverse velocity is $v = \partial y/\partial t = -i\omega y$. The characteristic impedance is defined as the ratio of force to velocity:
\[
Z_{\text{string}} = \frac{F}{v} = \frac{-T (ik y)}{-i\omega y} = \frac{Tk}{\omega}.
\]
Using the dispersion relation $\omega = \sqrt{T/\mu}\,k$, we obtain $Z_{\text{string}} = \sqrt{T\mu}$, a \textbf{real, positive constant} \cite{Morse1948}. A real impedance means that force and velocity are in phase, and energy is transported without dissipation.

\paragraph{Generalizing: Impedance from the Klein-Gordon Equation.}

Consider the massive Klein–Gordon equation in one dimension,
\begin{equation}
\frac{1}{c^2}\frac{\partial^2\psi}{\partial t^2} - \frac{\partial^2\psi}{\partial x^2} + \frac{m^2c^2}{\hbar^2}\psi = 0.
\label{eq:KG_1d_new}
\end{equation}
For a plane wave \(\psi(x,t) = \psi_0 e^{i(kx - \omega t)}\), the dispersion relation is \(\omega^2 = \omega_0^2 + c^2 k^2\) with \(\omega_0 = mc^2/\hbar\). By analogy with the string, we identify the spatial derivative \(-\partial\psi/\partial x\) as a generalized force and the temporal derivative \(\partial\psi/\partial t\) as a generalized velocity, leading to the wave impedance
\begin{equation}
Z(\omega) = \frac{-\partial\psi/\partial x}{\partial\psi/\partial t} = \frac{k}{\omega}.
\label{eq:Z_k_over_omega}
\end{equation}
Using the dispersion relation to eliminate \(k\) yields
\begin{equation}
Z(\omega) = \pm\frac{1}{c}\sqrt{1 - \frac{\omega_0^2}{\omega^2}},
\label{eq:impedance_final_new}
\end{equation}
where the sign corresponds to the propagation direction. Introducing a system-dependent constant \(Z_0\) (with dimensions of impedance) that encapsulates the specific physical context, we write
\begin{equation}
Z_{\text{phys}}(\omega) = Z_0 \, \sqrt{1 - \frac{\omega_0^2}{\omega^2}}.
\label{eq:impedance_physical}
\end{equation}
For the massless Klein–Gordon field, \(Z_0 = 1/c\); for electromagnetic waves in a cold plasma, \(Z_0 = \mu_0 c\). The same factor \(\sqrt{1-\omega_0^2/\omega^2}\) appears in the wave impedance of TE/TM modes in hollow waveguides \cite{Jackson1998,Pozar2011}, demonstrating that this mathematical form is far from arbitrary.

\begin{figure}[htbp]
\centering
\includegraphics[width=0.9\columnwidth]{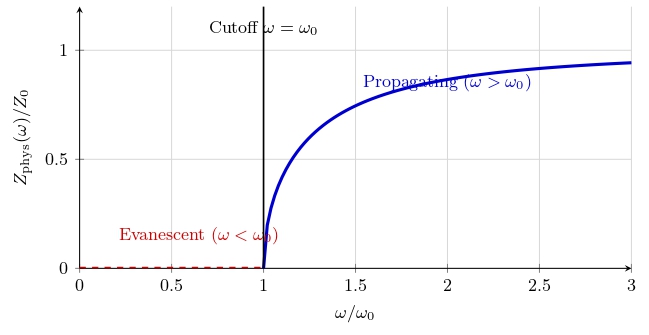}
\caption{Frequency-dependent wave impedance of a Klein–Gordon medium. For $\omega > \omega_0$, the impedance is real and positive, indicating propagating waves with energy transport. At $\omega = \omega_0$, the impedance drops to zero. For $\omega < \omega_0$, the wave number becomes imaginary and the impedance is purely imaginary (not shown), signalling evanescent decay.}
\label{fig:impedance_KG}
\end{figure}

The key features are:
\begin{itemize}
    \item \textbf{$\omega > \omega_0$ (propagating regime):} $Z$ is \textbf{real and positive}. Force and velocity are in phase, and energy flows.
    \item \textbf{$\omega = \omega_0$ (cutoff):} $Z = 0$. The medium oscillates uniformly with no spatial variation, and no energy is transmitted.
    \item \textbf{$\omega < \omega_0$ (evanescent regime):} $k = i\kappa$, giving $Z = i\kappa/\omega$, a \textbf{purely imaginary} impedance. Force and velocity are in quadrature, and the fields decay exponentially without net power flow.
\end{itemize}

\paragraph{Physical Examples.}

Table~\ref{tab:impedance_examples} summarizes several physical systems whose wave impedance follows Eq.~\eqref{eq:impedance_physical}. In each case, the impedance concept is rigorously defined within its own domain; the common factor \(\sqrt{1-\omega_0^2/\omega^2}\) reveals the underlying unity.

\begin{table*}[htbp]
\centering
\caption{Physical systems described by the Klein–Gordon dispersion and their wave impedance. The constant \(Z_0\) is system-specific; for the pure Klein–Gordon field, the impedance is an analogical extension.}
\label{tab:impedance_examples}
\begin{tabular}{|l|c|c|c|}
\hline
\textbf{System} & $\omega_0$ & $Z_0$ & $Z(\omega)/Z_0$ \\
\hline
Cold plasma (electromagnetic) \cite{Chen2016} & $\omega_p$ (plasma frequency) & $\mu_0 c$ & $\sqrt{1 - \omega_p^2/\omega^2}$ \\
\hline
Transmission line with shunt inductance \cite{Pozar2011} & $1/\sqrt{LC}$ (cutoff) & $\sqrt{L/C}$ & $\sqrt{1 - \omega_0^2/\omega^2}$ \\
\hline
Superconductor (London model) \cite{Tinkham2004} & $c/\lambda_L$ & $\mu_0 c$ & $\sqrt{1 - c^2/(\lambda_L^2\omega^2)}$ \\
\hline
Photonic crystal near band edge \cite{Joannopoulos2011} & $\omega_{\text{edge}}$ & $Z_{\text{eff}}$ (structure-dependent) & $\sqrt{1 - \omega_{\text{edge}}^2/\omega^2}$ \\
\hline
\end{tabular}
\end{table*}

\paragraph{Impedance and Reflection.}
At an interface between two media, the amplitude reflection coefficient is \(r = (Z_2 - Z_1)/(Z_2 + Z_1)\). This universal formula, derived solely from continuity conditions, allows us to predict total reflection when \(Z_2\) is imaginary, perfect matching when \(Z_1=Z_2\), and the existence of surface-bound states when \(Z_1+Z_2=0\) \cite{Jackson1998,Maier2007,Raether1988}.

Thus, even within the cautious interpretation advocated here, the wave impedance extracted from \(\omega(k)\) provides a unified language for discussing boundary phenomena across disciplines.

\section{Classical Analogies: Mass-Spring Chains and Brillouin Zones}

We now apply the interpretative framework developed in Section~2 to a purely classical system: the mass-spring chain. The same mathematical features we examined in the Klein-Gordon dispersion---slope, curvature, intercepts, and the ratio $k/\omega$---will reveal the same physical quantities: phase and group velocities, effective mass, density of states, and impedance. The conceptual insights gained from analyzing the Klein-Gordon dispersion extend naturally to such classical mechanical systems, providing tangible analogies that bridge quantum and classical physics. Among these classical analogies, the mass-spring chain stands out as particularly illuminating, revealing fundamental concepts of wave propagation in periodic structures through simple Newtonian mechanics.

Consider an infinite chain of identical masses $m$ connected by identical springs with spring constant $K$, with an equilibrium separation distance $a$ between adjacent masses. When the $n$th mass undergoes displacement $u_n$, Newton's second law yields
\begin{equation}
\begin{split}
m\frac{d^2u_n}{dt^2} &= K(u_{n-1} - u_n) + K(u_{n+1} - u_n) \\&= K(u_{n+1} + u_{n-1} - 2u_n).
\label{eq:mass_spring_eq}
\end{split}
\end{equation}
Seeking traveling wave solutions $u_n(t) = Ae^{i(kna - \omega t)}$ gives the dispersion relation
\begin{equation}
\omega^2 = \frac{2K}{m}(1 - \cos(ka)) = \frac{4K}{m}\sin^2\left(\frac{ka}{2}\right).
\label{eq:monatomic_dispersion}
\end{equation}

\begin{figure}[htbp]
\centering
\includegraphics[width=0.9\columnwidth]{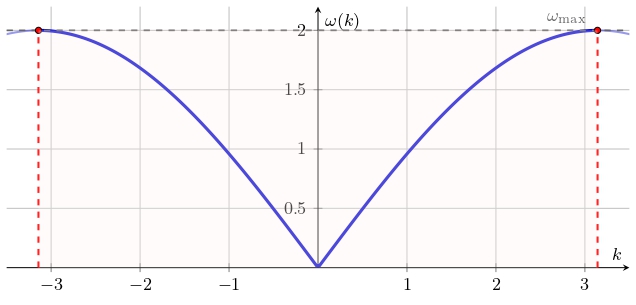}
\caption{Dispersion relation for a monatomic mass-spring chain: $\omega(k) = 2\sqrt{K/m}|\sin(ka/2)|$. The relation exhibits periodicity with period $2\pi/a$ in $k$-space, reflecting the discrete spatial periodicity $a$ of the chain. The red dashed lines demarcate the first Brillouin zone: all physically distinct wave numbers lie between $-\pi/a$ and $\pi/a$. At the zone boundaries $k = \pm \pi/a$, the dispersion curve flattens ($d\omega/dk = 0$), producing standing waves where neighboring masses move in perfect opposition. The maximum frequency $\omega_{\text{max}} = 2\sqrt{K/m}$ represents a cutoff beyond which no propagating waves exist. This simple mechanical system illustrates fundamental concepts including Brillouin zones, zone boundary behavior, and the relationship between spatial periodicity and $k$-space periodicity that underlies much of solid state physics and photonics.}
\label{fig:monatomic_chain}
\end{figure}

Equation \eqref{eq:monatomic_dispersion} contains remarkable insights about wave propagation in periodic systems, as visualized in Fig.~\ref{fig:monatomic_chain}. Most fundamentally, it exhibits periodicity in $k$-space with period $2\pi/a$. Increasing $k$ by $2\pi/a$ gives $e^{i(k+2\pi/a)na} = e^{ikna}e^{i2\pi n} = e^{ikna}$, so masses cannot distinguish wave numbers differing by $2\pi/a$. This leads directly to the concept of the Brillouin zone: only wave numbers between $-\pi/a$ and $\pi/a$ possess physical distinctness. This insight, first developed by Brillouin in his analysis of wave propagation in periodic structures \cite{Brillouin1953}, reappears throughout physics involving periodic media.

From this dispersion relation emerge the phase and group velocities that characterize wave propagation. The phase velocity $v_p = \omega/k = (2\sqrt{K/m}/k)|\sin(ka/2)|$ describes how wave crests appear to move, while the group velocity $v_g = d\omega/dk = a\sqrt{K/m}\cos(ka/2)$ describes how energy and information actually propagate. Near the zone center ($k \approx 0$), both velocities approach $a\sqrt{K/m}$, and the chain behaves as a continuum medium. At the zone boundary ($k = \pi/a$), however, $v_g$ vanishes completely while $v_p$ remains finite. This separation of velocities mirrors our Klein-Gordon analysis, where $v_p$ could exceed $c$ while $v_g$ remained subluminal.

The effective mass concept proves equally revealing here. Expanding $\omega(k)$ near the zone center gives $\omega \approx a\sqrt{K/m}|k|$, yielding infinite effective mass (since $d^2\omega/dk^2 = 0$). Near the zone boundary, however, expansion around $k = \pi/a$ gives $\omega \approx \omega_{\text{max}} - (a^2/8)\sqrt{K/m}(k - \pi/a)^2$, producing a negative effective mass $m_{\text{eff}} = -4m/(a^4K)$. This negative mass explains why waves near the zone boundary accelerate opposite to applied forces, a phenomenon observed in photonic crystals and semiconductor superlattices.

\begin{figure}[htbp]
\centering
\includegraphics[width=0.9\columnwidth]{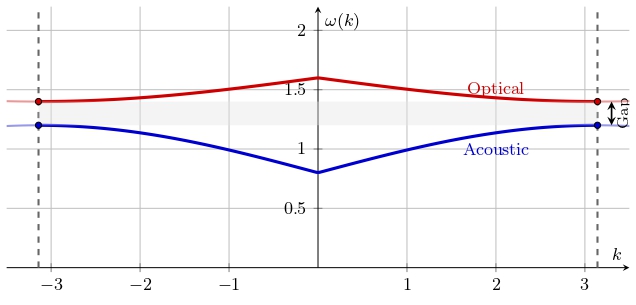}
\caption{Dispersion relation for a diatomic mass-spring chain with alternating masses $m_1$ and $m_2$ ($m_2 = 2m_1$). Two branches appear: a lower acoustic branch where masses move together at long wavelengths, and an upper optical branch where neighboring masses move oppositely even at $k=0$. A frequency gap (gray shaded region) opens between the branches, representing frequencies where no propagating wave solutions exist. The gap width $\Delta\omega$ increases with mass ratio $m_2/m_1$. At the zone boundary $k = \pi/a$, both branches flatten with $d\omega/dk = 0$, producing standing waves. This simple mechanical system illustrates fundamental concepts of solid state physics including acoustic and optical phonons, band gaps, and the relationship between structural periodicity and allowed frequency ranges.}
\label{fig:diatomic_chain}
\end{figure}

Now consider breaking the chain's perfect symmetry by alternating between two different masses $m_1$ and $m_2$. This diatomic chain models real crystals like NaCl, where sodium and chlorine atoms have different masses. The equations of motion now distinguish even and odd numbered sites:
\begin{equation}
\begin{split}
m_1\frac{d^2u_{2n}}{dt^2} &= K(u_{2n+1} + u_{2n-1} - 2u_{2n}), \\
m_2\frac{d^2u_{2n+1}}{dt^2} &= K(u_{2n+2} + u_{2n} - 2u_{2n+1}).
\end{split}
\end{equation}
Seeking solutions with the appropriate periodicity leads to the dispersion relation
\begin{equation}
\omega^2 = \frac{K(m_1 + m_2)}{m_1 m_2} \left[ 1 \pm \sqrt{1 - \frac{4m_1 m_2}{(m_1 + m_2)^2} \sin^2(ka/2)} \right],
\label{eq:diatomic_dispersion}
\end{equation}
as shown in Fig.~\ref{fig:diatomic_chain}. The plus sign gives the optical branch, where adjacent masses oscillate in opposite directions; the minus sign gives the acoustic branch, where they move together. These names originate from how these modes interact with light: optical modes create oscillating dipoles that couple to electromagnetic radiation, while acoustic modes do not \cite{Kittel2005}.

The emergence of two distinct branches follows from symmetry breaking. In the monatomic chain, all sites are equivalent, yielding one band. With two different masses, the unit cell doubles in size, halving the Brillouin zone and allowing two distinct oscillation patterns within each cell. Between the branches opens a frequency gap where no propagating solutions exist. This gap width $\Delta\omega$ grows with the mass ratio $m_2/m_1$, illustrating how increased asymmetry enhances band separation \cite{Born1912}.

\subsection{Density of States and van Hove Singularities}

The density of states for the monatomic chain follows from the same mode-counting logic developed for the Klein-Gordon system (see Appendix~\ref{app:dos_derivation}). In one dimension with length $L$, the number of modes with wave number between $k$ and $k+dk$ is $dN = (L/\pi)\,dk$, where the factor $1/\pi$ (rather than $1/2\pi$) accounts for both positive and negative $k$. Changing variables from $k$ to $\omega$ via $dN/d\omega = (L/\pi)\,dk/d\omega$ and using the dispersion relation~\eqref{eq:monatomic_dispersion} yields
\begin{equation}
g(\omega) = \frac{L}{\pi}\frac{dk}{d\omega} = \frac{L}{\pi a}\frac{1}{\sqrt{\omega_{\text{max}}^2 - \omega^2}},
\label{eq:dos_monatomic}
\end{equation}
with $\omega_{\text{max}} = 2\sqrt{K/m}$. This density of states exhibits a van Hove singularity at $\omega = \omega_{\text{max}}$, where $d\omega/dk = 0$ \cite{VanHove1953}. This singularity, visible as the flat region at the Brillouin zone boundary in Fig.~\ref{fig:monatomic_chain}, has measurable consequences. In the specific heat of solids at low temperatures, the density of states of acoustic phonons near the Brillouin zone boundary determines the characteristic $T^3$ temperature dependence observed in insulators \cite{Debye1912}.

The pedagogical power of mass-spring chains lies in how they embody abstract concepts in concrete mathematics. Beyond propagation characteristics, these chains also exhibit impedance phenomena directly analogous to the Klein-Gordon case. The impedance for the monatomic chain, defined as the ratio of force to velocity at a drive point, takes the form $Z(\omega) \propto \sqrt{Km}/\sqrt{1 - (\omega/\omega_{\text{max}})^2}$ for $\omega < \omega_{\text{max}}$, diverging at the maximum frequency $\omega_{\text{max}}$ just as the Klein-Gordon impedance diverges at $\omega_0$. This impedance governs how vibrations reflect at boundaries or couple between chain segments with different properties. The distinction between phase velocity (wave crest motion) and group velocity (energy propagation) emerges clearly from the mathematics, with $v_g$ vanishing at zone boundaries while $v_p$ remains finite. From this simple system, students learn universal principles that transfer to electrons in crystals, light in photonic materials, and sound in acoustic metamaterials \cite{BlochMetamaterial}: spatial periodicity produces $k$-space periodicity (Brillouin zones); broken symmetry opens band gaps; flat dispersion creates high density of states; zone boundaries host standing waves with vanishing group velocity; effective mass emerges from dispersion curvature; and impedance controls boundary interactions.

\section{Hydrodynamic Waves: Gravity and Surface Tension in Competition}

The same five questions we asked of the Klein–Gordon dispersion---about propagation speeds, forbidden frequencies, available modes, inertial response, and boundary behavior---can be asked of water waves. Here, the dispersion relation changes form entirely, yet the extraction procedure remains identical. Water waves provide another classical system where dispersion relations reveal beautifully the competition between different physical restoring mechanisms, offering observable phenomena that students can analyze using the interpretative framework developed for the Klein–Gordon equation.

When a stone disturbs a pond's surface, careful observation reveals two distinct wave types engaged in dynamical competition. The large, leisurely waves that spread broadly represent gravity's influence. The tiny, rapid ripples that race ahead belong to surface tension. Between these extremes exist waves that cannot decide which parent to obey, displaying minimal propagation speeds. This natural drama on water's surface is captured by one of physics' most elegant dispersion relations \cite{Craik2004}.

\subsection{The Dispersion Relation: Deep Water, Shallow Water, and the Role of Depth}

For a first approach, consider the limit where the water depth \(h\) is much larger than the wavelength \(\lambda\) (i.e., \(kh \gg 1\), with \(k = 2\pi/\lambda\)). In this \textbf{deep‑water} regime the dispersion relation takes the compact form
\begin{equation}
\omega^2 = gk + \frac{\sigma}{\rho} k^3,
\label{eq:water_dispersion}
\end{equation}
where \(g \approx 9.8\ \text{m/s}^2\) is the gravitational acceleration, \(\sigma \approx 0.073\ \text{N/m}\) is the surface tension, and \(\rho \approx 10^3\ \text{kg/m}^3\) is the density of water \cite{Whitham1974,Lighthill1978}. Equation~\eqref{eq:water_dispersion} embodies a literal physical tug‑of‑war: the term \(gk\) declares ``gravity matters,'' the term \((\sigma/\rho)k^3\) declares ``surface tension matters,'' and which force dominates depends entirely on the wave's spatial scale \cite{OndasGravidade}.

Why does spatial scale determine the dominant restoring force? Consider a long ocean swell with a wavelength of hundreds of metres. Lifting such a vast water column requires working against gravity's substantial weight, whereas the gentle surface curvature involves negligible surface‑tension energy. Now imagine a tiny ripple with a millimetre‑scale wavelength: elevating that thin ribbon of water poses little challenge for gravity, but sharply bending the surface demands considerable surface‑tension energy. This scale dependence is analogous to the difference between bending a long wooden plank (gravity resists strongly) and crumpling a sheet of paper (surface stiffness resists strongly) \cite{Whitham1974}. The generation of these wave fields by wind is a rich topic in its own right \cite{Phillips1957,Phillips1958}.

Equation~\eqref{eq:water_dispersion} is a deep‑water approximation. The full dispersion relation for a fluid layer of finite depth \(h\) reads
\begin{equation}
\omega^2 = \Bigl(gk + \frac{\sigma}{\rho}k^3\Bigr)\,\tanh(kh).
\label{eq:full_water}
\end{equation}
The factor \(\tanh(kh)\) interpolates between two extreme regimes:
\begin{itemize}
  \item \textbf{Deep water} (\(kh \gg 1\)): \(\tanh(kh) \approx 1\), and we recover Eq.~\eqref{eq:water_dispersion}. Waves are dispersive; phase and group velocities differ.
  \item \textbf{Shallow water} (\(kh \ll 1\)): \(\tanh(kh) \approx kh\). If, in addition, surface tension is negligible (\(\sigma \to 0\)), Eq.~\eqref{eq:full_water} reduces to \(\omega^2 = gh\,k^2\), a \textbf{linear, non‑dispersive} relation with \(v_p = v_g = \sqrt{gh}\).
\end{itemize}
The shallow‑water regime is not a curiosity: tsunamis, with wavelengths \(\lambda \sim 100\ \text{km}\) far exceeding the typical ocean depth \(h \sim 4\ \text{km}\), satisfy \(kh \ll 1\). Consequently, their phase and group velocities are both equal to \(\sqrt{gh} \approx 200\ \text{m/s}\) (about \(720\ \text{km/h}\)), and the entire wave packet travels at a single speed without spreading \cite{NOAA_tsunami}. This non‑dispersive character, not the \(\omega \propto \sqrt{k}\) behaviour of deep‑water gravity waves, is what makes tsunami early‑warning systems effective.

In what follows we concentrate on the deep‑water regime, Eq.~\eqref{eq:water_dispersion}, because it is here that the competition between gravity and surface tension becomes most transparent, and it is this regime that wind‑generated swells and ripples inhabit. The interpretation tools we develop, however, can be applied equally well to the full relation~\eqref{eq:full_water}.

\subsection{Phase and Group Velocities}

The phase velocity derived from Eq.~\eqref{eq:water_dispersion},
\begin{equation}
v_p = \frac{\omega}{k} = \sqrt{\frac{g}{k} + \frac{\sigma k}{\rho}},
\end{equation}
narrates a compelling physical story. For long waves (small \(k\)), \(v_p \approx \sqrt{g/k}\): longer waves travel faster! This explains why long‑wavelength swells race ahead of shorter waves on the open ocean. For short ripples (large \(k\)), \(v_p \approx \sqrt{\sigma k/\rho}\): tinier waves also travel faster! The intriguing consequence is a minimum phase velocity at \(k = \sqrt{\rho g/\sigma}\). For water at standard conditions this corresponds to a wavelength \(\lambda \approx 1.7\ \text{cm}\), where waves crawl at merely \(23\ \text{cm/s}\). Observing a pond reveals these intermediate ripples as the slowest waves, being overtaken by both larger gravity waves and smaller capillary dimples \cite{OndasGravidade}.

Now consider the group velocity, the true carrier of energy and information:
\begin{equation}
v_g = \frac{d\omega}{dk} = \frac{1}{2\omega}\Bigl(g + \frac{3\sigma k^2}{\rho}\Bigr).
\end{equation}
For pure gravity waves (\(\sigma = 0\)), \(v_g = v_p/2\): wave crests outpace their own energy by a factor of two. For pure capillary waves (\(g = 0\)), \(v_g = 3v_p/2\): energy now leads the crests. This dramatic difference---and the reversal of which is larger---explains the intricate V‑shaped patterns of ship wakes: the wake angle derives from \(v_g/v_p\), with gravity waves forming the outer arms and capillary waves creating the inner feathering \cite{Whitham1974,Mei1989}. Students can observe this directly: a swimming duck produces both gravity waves (larger, slower) and capillary waves (smaller, faster) with distinct propagation characteristics.

\subsection{Effective Mass and Curvature: When Waves ``Remember'' Their Past}
\label{sec:water_effective_mass}

While the term \textit{effective mass} originates in solid‑state physics, where it quantifies the inertial response of electrons in a periodic potential, the underlying mathematics—the inverse of the dispersion curvature \(d^2\omega/dk^2\)—governs wave‑packet dynamics in any dispersive medium. In what follows, we borrow the language of effective mass as an analogy: a wave packet in a region of negative curvature behaves \textit{as if} it possessed a negative effective mass, meaning its spreading is counteracted by the dispersion itself. This should not be mistaken for a literal attribution of mechanical inertia to a fluid element; it is a concise way of describing how the shape of \(\omega(k)\) controls the fate of a wave packet.

\begin{figure}[htbp]
\centering
\includegraphics[width=\columnwidth]{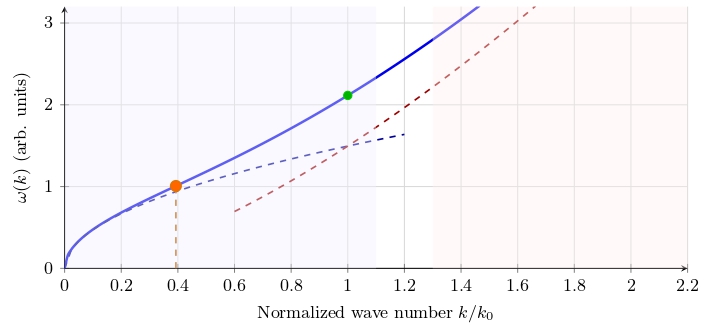}
\includegraphics[width=\columnwidth]{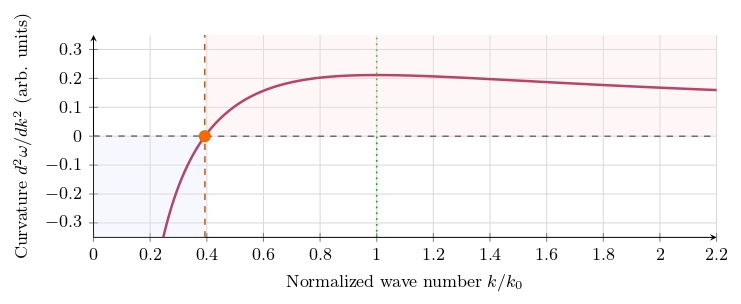}
\caption{Top: Dispersion relation $\omega^2 = gk + (\sigma/\rho)k^3$ for deep-water waves, plotted against normalized wave number $k/k_0$ with $k_0 = \sqrt{\rho g/\sigma}$. The blue dashed curve shows the pure gravity asymptote; the red dashed curve shows the pure capillary asymptote. The green point marks the minimum of the phase velocity at $k=k_0$ ($k/k_0=1$). The orange point marks the inflection point of the dispersion curve at $k_{\mathrm{infl}} \approx 0.393\,k_0$. Shaded regions indicate gravity-dominated (blue) and capillary-dominated (red) regimes. Bottom: Second derivative $d^2\omega/dk^2$ showing the transition from negative curvature (negative effective mass, blue region) to positive curvature (positive effective mass, red region). The curvature vanishes at $k_{\mathrm{infl}}$, where wave packets propagate with minimal distortion.}
\label{fig:curvature_water}
\end{figure}

The mathematical story begins with the exact curvature from our dispersion relation (Eq.~\eqref{eq:water_dispersion}):
\begin{equation}
\begin{split}
\frac{d^2\omega}{dk^2} = \frac{6g\frac{\sigma}{\rho}k^2 + 3\left(\frac{\sigma}{\rho}\right)^2 k^4 - g^2}{4\omega^3}.
\label{eq:exact_curvature}
\end{split}
\end{equation}
This curvature behavior, visualized in Fig.~\ref{fig:curvature_water}, reveals a remarkable feature: the second derivative changes sign at a specific wave number, marking a transition between fundamentally different physical regimes. The transition is governed by the characteristic wave number
\begin{equation}
k_0 \equiv \sqrt{\frac{\rho g}{\sigma}},
\label{eq:k0_definition}
\end{equation}
which corresponds to the minimum of the phase velocity. For water at room temperature, \(k_0 \approx 366\ \text{rad/m}\) (\(\lambda_0 = 2\pi/k_0 \approx 1.7\ \text{cm}\)). With this scale in hand, we can read the conversation between gravity and surface tension directly from the sign of the curvature.

In the gravity-dominated world of long waves (\(k \ll k_0\)), the equation simplifies to a whisper:
\begin{equation}
\frac{d^2\omega}{dk^2} \approx -\frac{\sqrt{g}}{4}k^{-3/2} < 0.
\end{equation}
That minus sign is the surprise: negative curvature means that, in the effective‑mass analogy, the wave packet behaves as if it had negative mass. This is ``anti‑inertia'': the dispersion relation acts to hold the wave packet together rather than spread it apart. This explains why long deep‑water swells can travel vast distances across oceans without losing their shape as quickly as one might expect---their negative effective mass counteracts the usual tendency of waves to disperse \cite{Whitham1974}.

Now visit the tiny world of capillary waves (\(k \gg k_0\)), where surface tension rules:
\begin{equation}
\frac{d^2\omega}{dk^2} \approx \frac{3}{4}\sqrt{\frac{\sigma}{\rho}}k^{-1/2} > 0.
\end{equation}
Here the curvature is positive, giving ordinary positive effective mass. These waves disperse rapidly: blow on your tea and watch how quickly the ripples scatter in all directions, losing their initial pattern almost immediately.

Between these extremes lies a special scale, the inflection point where \(d^2\omega/dk^2 = 0\). Solving the exact curvature equation yields
\begin{equation}
k_{\text{infl}} = \sqrt{\frac{2\sqrt{3}-3}{3}}\,k_0 \approx 0.393\,k_0,
\label{eq:kinfl_correct}
\end{equation}
corresponding to a wavelength of approximately \(4.4\ \text{cm}\) for water. As shown in Fig.~\ref{fig:curvature_water} (bottom panel), the curvature vanishes at this point, meaning the effective mass diverges. At this precise scale, the group velocity dispersion changes sign, and wave packets experience minimal spreading during propagation---the competing effects of gravity and surface tension nearly cancel. In a carefully controlled wave tank, one can observe that waves of this specific wavelength maintain their shape over distances noticeably longer than either longer gravity waves or shorter capillary ripples, providing a striking demonstration of dispersion management through scale selection.

This curvature analysis connects beautifully to our earlier discussions. As Dias and Kharif note, ``The gravity-capillary transition provides a classical laboratory for phenomena typically associated with quantum systems'' \cite{Dias1999}. Indeed, the same mathematics that describes electrons in crystals or photons in periodic structures here describes ripples on a pond. The effective mass isn't just an abstract concept; it's the reason why some waves remember their past (long swells crossing oceans) while others live only in the moment (capillary ripples vanishing in seconds).

For students, the lesson is wonderfully tangible. Throw a big stone: watch the gravity waves maintain their circular pattern. Drop a tiny pebble: see the capillary waves scatter immediately. And if you're patient and precise, create waves of about \(4\ \text{cm}\) wavelength: marvel at how they keep their shape. All these behaviors live in Eq.~\eqref{eq:water_dispersion}, waiting to be discovered by those who know how to listen to what curvature has to say. The visual representation in Fig.~\ref{fig:curvature_water} encapsulates this multilayered interpretation: the top panel shows how \(\omega(k)\) smoothly transitions between asymptotic behaviors (blue and red dashed curves), while the bottom panel reveals the sign change in curvature that governs wave packet dynamics across scales.

\subsection{Density of States: How Many Waves Fit in a Frequency Interval?}

The density of states \(g(\omega)\) answers a simple question: given a frequency \(\omega\), how many independent wave modes exist per unit frequency? The concept is exactly the same as for the Klein–Gordon system (Section~2.3), but now the dispersion relation is \(\omega^2 = gk + (\sigma/\rho)k^3\) and the waves live on a two‑dimensional surface.

\begin{figure}[htbp]
\centering
\includegraphics[width=\columnwidth]{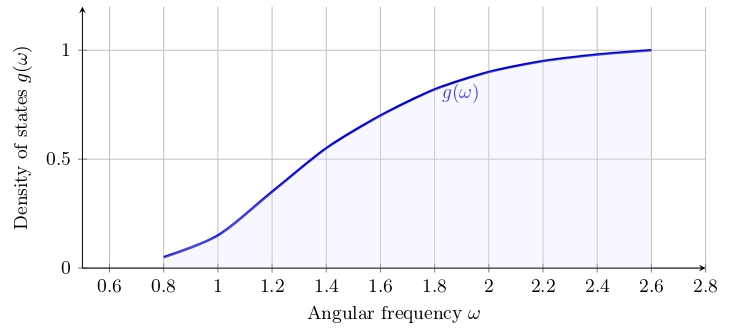}
\caption{Density of states \(g(\omega)\) for deep‑water waves (schematic). The curve increases monotonically with frequency: at low \(\omega\) (gravity‑dominated waves) the density grows slowly; at high \(\omega\) (capillary‑dominated waves) it rises more steeply. The shaded area represents the total number of modes up to a given frequency.}
\label{fig:density_states_water}
\end{figure}

For a surface of area \(A\), the number of modes with wave number between \(k\) and \(k+dk\) is \(dN = (A/2\pi)\,k\,dk\) \cite{Whitham1974,Lighthill1978}. Changing variable from \(k\) to \(\omega\) using the dispersion relation yields the density of states in parametric form:
\[
g(\omega(k)) = \frac{A}{\pi}\,
\frac{k\,\omega(k)}{g + 3\frac{\sigma}{\rho}k^{2}},
\qquad
\omega(k) = \sqrt{gk + \frac{\sigma}{\rho}k^{3}}.
\]
This expression is plotted in Fig.~\ref{fig:density_states_water}. It tells us how the available modes are distributed across frequencies.

The curve increases monotonically: the higher the frequency, the more wave modes are available. This happens because shorter waves (larger \(k\)) pack more densely in \(k\)‑space. At low frequencies, where gravity dominates, the density grows relatively slowly; at high frequencies, where surface tension takes over, the growth becomes steeper. The smooth bend in the curve reflects the transition between the two regimes---exactly the same interplay of \(gk\) and \((\sigma/\rho)k^3\) that we have been studying throughout this section.

Notice that unlike the Klein–Gordon case, there is no gap: waves exist at all frequencies. The difference lies in the absence of a minimum frequency (no \(\omega_0\) term). The deep‑water dispersion relation allows arbitrarily slow waves (very long gravity waves) because \(\omega \to 0\) as \(k \to 0\). Consequently, \(g(\omega)\) is non‑zero down to \(\omega = 0\).

In summary, the density of states translates the geometry of the dispersion curve into a practical tool: it tells us how the system’s vibrational “menu” is organized. Although we have not derived the formula here (the mode‑counting procedure is identical in spirit to that in Appendix~\ref{app:dos_derivation}), its physical content is simple to grasp once \(\omega(k)\) is known.

\section{Conclusion: Reading Physics from Mathematical Forms}

This paper has pursued a simple pedagogical conviction: that the mathematical form of a dispersion relation is already a physical narrative, waiting to be read. From the single curve \(\omega(k)\) we extracted five universal quantities---phase velocity, group velocity, density of states, effective mass, and impedance---each tied to a specific mathematical feature of that curve. The Klein--Gordon form \(\omega^2 = \omega_0^2 + c^2k^2\) served as our prototype, but the extraction procedure proved equally effective for classical mass-spring chains and for water waves governed by the interplay of gravity and surface tension.

The skill we hope students will acquire is a systematic questioning mindset. Faced with an unfamiliar \(\omega(k)\), they should ask: How do disturbances propagate? Which frequencies are forbidden? How many modes are available? How does the medium respond to forces? How do waves behave at boundaries? These questions are not tied to any specific subfield---they are the grammar of wave physics, applicable from quantum field theory to oceanography.

A crucial lesson, brought home by the hydrodynamic case, concerns the domain of validity. The deep-water dispersion \(\omega^2 = gk + (\sigma/\rho)k^3\) vividly captures the gravity-capillary competition, yet it is an approximation. Including the finite-depth factor \(\tanh(kh)\), as discussed in Section~4.2, reveals a shallow-water regime where waves become non-dispersive---the regime that governs tsunamis. Recognizing which terms dominate in a given physical regime is as important as extracting physics from the formula itself.

Ultimately, what we have called ``dispersion relation literacy'' is the ability to look at a plot of \(\omega\) versus \(k\) and see not merely a graph but a map: of velocities, gaps, available states, inertia, and resistance. Across quantum fields, elastic chains, and ocean surfaces, the language of dispersion remains the same. By placing interpretation on an equal footing with derivation, we give students a tool that remains useful long after specific equations have been forgotten---a way of reading the physical world through the shapes that describe its waves.

\appendix

\section{Detailed Derivation of the Density of States}
\label{app:dos_derivation}

We derive $g(\omega)$ for the Klein-Gordon dispersion $\omega^2 = \omega_0^2 + c^2k^2$ in three dimensions. The logic proceeds in four steps: (i) discretize $\mathbf{k}$-space via periodic boundaries, (ii) count modes in a spherical shell, (iii) change variables from $k$ to $\omega$, (iv) interpret the result.

\paragraph{Step 1: Discretizing $\mathbf{k}$-space.}

Consider a cube of side $L$ and volume $V = L^3$. Imposing periodic boundary conditions $\psi(x+L,y,z,t) = \psi(x,y,z,t)$ (and similarly for $y,z$) forces the wave vector components to satisfy:
\begin{equation}
e^{i(k_x L)} = 1 \;\Longrightarrow\; k_x L = 2\pi n_x,\quad n_x \in \mathbb{Z},
\end{equation}
and analogously for $k_y, k_z$. Hence:
\begin{equation}
\mathbf{k} = \frac{2\pi}{L}(n_x, n_y, n_z), \qquad n_x, n_y, n_z \in \mathbb{Z}.
\label{eq:discrete_k}
\end{equation}
The factor $2\pi/L$ is the spacing between neighboring $\mathbf{k}$-points. Each distinct triple $(n_x,n_y,n_z)$ corresponds to one independent plane wave mode $e^{i(\mathbf{k}\cdot\mathbf{r}-\omega t)}$.

\paragraph{Step 2: Number of modes in a spherical shell.}

In $\mathbf{k}$-space, each mode occupies a volume $(2\pi/L)^3 = (2\pi)^3/V$. Therefore, the number of modes in a small volume $d^3k$ is:
\begin{equation}
dN = \frac{d^3k}{(2\pi)^3/V} = \frac{V}{(2\pi)^3}\, d^3k.
\label{eq:mode_density}
\end{equation}
Because $\omega$ depends only on $k = |\mathbf{k}|$ (isotropy), we integrate over spherical coordinates: $d^3k = 4\pi k^2 dk$. The number of modes with magnitude between $k$ and $k+dk$ is:
\begin{equation}
dN = \frac{V}{(2\pi)^3} \cdot 4\pi k^2 dk = \frac{V}{2\pi^2}\, k^2 dk.
\label{eq:dN_k}
\end{equation}
The factor $4\pi$ (full solid angle) is correct because $\mathbf{k}$ and $-\mathbf{k}$ represent distinct modes (waves traveling in opposite directions).

\paragraph{Step 3: Transforming from $k$ to $\omega$.}

We need $g(\omega) = dN/d\omega$. From Eq.~\eqref{eq:dN_k}:
\begin{equation}
\frac{dN}{dk} = \frac{V}{2\pi^2}\, k^2.
\label{eq:dN_dk}
\end{equation}
The dispersion relation $\omega^2 = \omega_0^2 + c^2k^2$ gives, by implicit differentiation:
\begin{equation}
2\omega = 2c^2 k \frac{dk}{d\omega} \;\Longrightarrow\; \frac{dk}{d\omega} = \frac{\omega}{c^2 k}.
\label{eq:dk_domega}
\end{equation}
This holds for $\omega > \omega_0$, where $k$ is real. Using the chain rule:
\begin{equation}
g(\omega) = \frac{dN}{d\omega} = \frac{dN}{dk} \cdot \frac{dk}{d\omega} = \frac{V}{2\pi^2} k^2 \cdot \frac{\omega}{c^2 k} = \frac{V}{2\pi^2 c^2}\, k\omega.
\label{eq:g_intermediate}
\end{equation}
Now eliminate $k$ using $k = \frac{1}{c}\sqrt{\omega^2 - \omega_0^2}$ (solving the dispersion relation for $k$):
\begin{equation}
\begin{split}
g(\omega) &= \frac{V}{2\pi^2 c^2} \cdot \frac{1}{c}\sqrt{\omega^2 - \omega_0^2} \cdot \omega \\&= \frac{V}{2\pi^2 c^3}\, \omega\sqrt{\omega^2 - \omega_0^2}, \qquad \omega > \omega_0.
\label{eq:g_final}
\end{split}
\end{equation}
For $\omega < \omega_0$, the wave number is imaginary ($k = i\kappa$), corresponding to evanescent (non-propagating) fields. These do not count as propagating modes, so:
\begin{equation}
g(\omega) = 0, \qquad \omega < \omega_0.
\label{eq:g_zero}
\end{equation}

\paragraph{Step 4: Checking the result.}

\textit{Dimensions:} $V/c^3$ has dimensions of $[\text{time}]^3$; $\omega$ and $\sqrt{\omega^2 - \omega_0^2}$ each have $[\text{time}]^{-1}$; product $g(\omega)$ has $[\text{time}]^{-1}$ as expected (number per unit frequency).

\textit{Limits:}
\begin{itemize}
\item $\omega_0 \to 0$ (massless): $g(\omega) = \dfrac{V}{2\pi^2 c^3}\,\omega^2$, the standard photon density of states.
\item $\omega \to \omega_0^+$: writing $\omega = \omega_0 + \varepsilon$, $\sqrt{\omega^2 - \omega_0^2} \approx \sqrt{2\omega_0\varepsilon}$, so $g(\omega) \propto \sqrt{\omega - \omega_0}$ (soft onset).
\item $\omega \gg \omega_0$: $g(\omega) \approx \dfrac{V}{2\pi^2 c^3}\,\omega^2$, recovering the massless behavior.
\end{itemize}

\paragraph{Step 5: Cumulative number of modes (optional).}

Integrating $g(\omega)$ from $\omega_0$ to $\omega$ gives the total number of modes up to frequency $\omega$:
\begin{equation}
\begin{split}
N(\omega) &= \int_{\omega_0}^{\omega} g(\omega')\,d\omega' = \frac{V}{2\pi^2 c^3} \int_{\omega_0}^{\omega} \omega'\sqrt{\omega'^2 - \omega_0^2}\,d\omega'\\&=\frac{V}{6\pi^2 c^3} (\omega^2 - \omega_0^2)^{3/2}.
\end{split}
\end{equation}
For $\omega_0 = 0$, this reduces to $N(\omega) = V\omega^3/(6\pi^2 c^3)$, the well-known result for photons.

\paragraph{Why this derivation matters.}

Every factor has a clear physical origin:
\begin{itemize}
\item $2\pi/L$: discretization from periodic boundaries.
\item $V/(2\pi)^3$: density of modes in $\mathbf{k}$-space.
\item $4\pi k^2 dk$: spherical shell integration over all directions.
\item $dk/d\omega = \omega/(c^2k)$: geometry of the dispersion curve.
\item $\sqrt{\omega^2 - \omega_0^2}$: the gap structure encoded in the dispersion.
\end{itemize}
This procedure is universal: for any isotropic dispersion $\omega(k)$, simply compute $k^2 (dk/d\omega)$ and multiply by $V/(2\pi^2)$. Historically, the density of states concept emerged from the study of blackbody radiation \cite{Planck1901,Jeans1905}, and later became a cornerstone of solid‑state physics \cite{Debye1912,Reif1965}.

\section{Catalog of Wave Equations and Dispersion Relations}
\label{sec:appendix}

To demonstrate the universality of our interpretative framework, Table~\ref{tab:dispersion_catalog} presents a catalog of fundamental wave equations and their dispersion relations. This compilation shows how identical mathematical forms recur across physics subdisciplines, from quantum mechanics to hydrodynamics, while encoding distinct physical mechanisms through parameter interpretation, establishing dispersion relations as a unifying language for wave phenomena.

Table~\ref{tab:symbols_glossary} lists the physical parameters used throughout the catalog.

\begin{table*}[ht]
\caption{Physical parameters appearing in the dispersion relations catalog.}
\label{tab:symbols_glossary}
\centering
\small
\renewcommand{\arraystretch}{1.2}
\begin{tabular}{@{}cl@{\hspace{1.5cm}}cl@{\hspace{1.5cm}}cl@{}}
\toprule
\textbf{Symbol} & \textbf{Physical Meaning} & \textbf{Symbol} & \textbf{Physical Meaning} & \textbf{Symbol} & \textbf{Physical Meaning} \\
\midrule
$\omega$ & Angular frequency & $k$ & Wave number & $c$ & Wave speed \\
$v_p = \omega/k$ & Phase velocity & $v_g = d\omega/dk$ & Group velocity & $\lambda$ & Wavelength \\
\hline
$m$ & Mass (particle) & $\hbar$ & Planck constant & $D$ & Diffusion coefficient \\
$g$ & Gravity & $\rho$ & Density & $\sigma$ & Surface tension \\
$h$ & Fluid depth & $K$ & Spring constant & $a$ & Lattice spacing \\
$E$ & Young's modulus & $I$ & Area moment & $A$ & Cross-section area \\
$L, C$ & Inductance, Capacitance & $R, G$ & Resistance, Conductance & $\alpha,\beta,\gamma$ & Parameters \\
\bottomrule
\end{tabular}
\end{table*}

\begin{table*}[ht]
\caption{Summary of fundamental wave equations and their dispersion relations. The functional form $\omega(k)$ encodes the essential physics of wave propagation in each system.}
\label{tab:dispersion_catalog}
\centering
\footnotesize
\renewcommand{\arraystretch}{1.4}

\noindent
\parbox{0.23\textwidth}{\centering\textbf{Equation (Name)}}
\hfill
\parbox{0.23\textwidth}{\centering\textbf{Dispersion Relation $\omega(k)$}}
\hfill
\parbox{0.23\textwidth}{\centering\textbf{Geometrical Behavior}}
\hfill
\parbox{0.23\textwidth}{\centering\textbf{Physical Interpretation}}

\vspace{2pt}
\hrule
\vspace{4pt}

\begin{minipage}{\textwidth}

\noindent
\parbox{0.23\textwidth}{
\textbf{Linear Wave (1D)}\\
$\displaystyle \frac{1}{c^2}\psi_{tt} = \psi_{xx}$\\
\cite{Whitham1974}
}
\hfill
\parbox{0.23\textwidth}{
$\omega^2 = c^2 k^2$\\
$\omega = \pm ck$
}
\hfill
\parbox{0.23\textwidth}{
Linear\\
$v_p = v_g = c$
}
\hfill
\parbox{0.23\textwidth}{
Non-dispersive\\
EM in vacuum, sound
}
\vspace{3pt}

\hrule
\vspace{3pt}

\noindent
\parbox{0.23\textwidth}{
\textbf{Free Schrödinger}\\
$i\hbar\psi_t = -\frac{\hbar^2}{2m}\psi_{xx}$\\
\cite{Messiah1961}
}
\hfill
\parbox{0.23\textwidth}{
$\omega = \frac{\hbar k^2}{2m}$
}
\hfill
\parbox{0.23\textwidth}{
Parabolic\\
$v_g = 2v_p$
}
\hfill
\parbox{0.23\textwidth}{
Matter waves\\
Normal dispersion
}
\vspace{3pt}

\hrule
\vspace{3pt}

\noindent
\parbox{0.23\textwidth}{
\textbf{Klein-Gordon}\\
$\frac{1}{c^2}\psi_{tt} = \psi_{xx} - \frac{m^2c^2}{\hbar^2}\psi$\\
\cite{Bjorken1965}
}
\hfill
\parbox{0.23\textwidth}{
$\omega^2 = c^2k^2 + \frac{m^2c^4}{\hbar^2}$\\
$\omega_0 = mc^2/\hbar$
}
\hfill
\parbox{0.23\textwidth}{
Hyperbolic\\
$v_p > c > v_g$
}
\hfill
\parbox{0.23\textwidth}{
Relativistic fields\\
Plasmas, superconductors
}
\vspace{3pt}

\hrule
\vspace{3pt}

\noindent
\parbox{0.23\textwidth}{
\textbf{Telegraph}\\
$\psi_{xx} = LC\psi_{tt} + RC\psi_t$\\
\cite{Heaviside1893}
}
\hfill
\parbox{0.23\textwidth}{
$k^2 = LC\omega^2 - iRC\omega$
}
\hfill
\parbox{0.23\textwidth}{
Complex $\omega(k)$\\
Attenuation
}
\hfill
\parbox{0.23\textwidth}{
Lossy lines\\
Damped waves
}
\vspace{3pt}

\hrule
\vspace{3pt}

\noindent
\parbox{0.23\textwidth}{
\textbf{Euler-Bernoulli}\\
$\psi_{tt} = -\frac{EI}{\rho A}\psi_{xxxx}$\\
\cite{LandauLifshitz1986}
}
\hfill
\parbox{0.23\textwidth}{
$\omega^2 = \frac{EI}{\rho A} k^4$\\
$\omega \propto k^2$
}
\hfill
\parbox{0.23\textwidth}{
$\omega \propto k^2$\\
Strong dispersion
}
\hfill
\parbox{0.23\textwidth}{
Flexural waves\\
Beam vibrations
}
\vspace{3pt}

\hrule
\vspace{3pt}

\noindent
\parbox{0.23\textwidth}{
\textbf{Deep Water}\\
(gravity)\\
\cite{Lamb1932}
}
\hfill
\parbox{0.23\textwidth}{
$\omega^2 = gk$
}
\hfill
\parbox{0.23\textwidth}{
$\omega \propto \sqrt{k}$\\
$v_g = v_p/2$
}
\hfill
\parbox{0.23\textwidth}{
Ocean swells\\
Longer $\lambda$ faster
}
\vspace{3pt}

\hrule
\vspace{3pt}

\noindent
\parbox{0.23\textwidth}{
\textbf{Capillary}\\
(surface tension)\\
\cite{Levich1962}
}
\hfill
\parbox{0.23\textwidth}{
$\omega^2 = \frac{\sigma}{\rho}k^3$
}
\hfill
\parbox{0.23\textwidth}{
$\omega \propto k^{3/2}$\\
$v_g = 3v_p/2$
}
\hfill
\parbox{0.23\textwidth}{
Small ripples\\
mm scale
}
\vspace{3pt}

\hrule
\vspace{3pt}

\noindent
\parbox{0.23\textwidth}{
\textbf{Water Waves}\\
(gravity + tension)\\
depth $h$\\
\cite{Phillips1977}
}
\hfill
\parbox{0.23\textwidth}{
$\omega^2 = \left(gk + \frac{\sigma}{\rho}k^3\right)\tanh(kh)$
}
\hfill
\parbox{0.23\textwidth}{
\vspace{0.1cm}
\textit{Deep water ($kh \gg 1$):}\\
$\tanh(kh)\approx 1$\\
$\omega^2 \approx gk + \frac{\sigma}{\rho}k^3$\\
$v_g \neq v_p$ (dispersive)\\
\vspace{0.2cm}
\textit{Shallow water ($kh \ll 1$):}\\
$\tanh(kh)\approx kh$\\
$\omega^2 \approx \left(gk + \frac{\sigma}{\rho}k^3\right)kh$\\
For $\sigma \to 0$: $\omega^2 = gh\,k^2$\\
$v_p = v_g = \sqrt{gh}$ (non-dispersive)
}
\hfill
\parbox{0.23\textwidth}{
\vspace{0.1cm}
General surface waves\\
Gravity-capillary competition\\
$\lambda_{\text{min}} \approx 1.7$ cm at $v_p$ min\\
\textit{Shallow water:} tsunamis, tides\\
$v_p$ independent of $k$, depends only on $h$
}
\vspace{3pt}
\hrule
\vspace{3pt}

\noindent
\parbox{0.23\textwidth}{
\textbf{Monatomic Chain}\\
$m\ddot{u}_n = K(u_{n+1}+u_{n-1}-2u_n)$\\
\cite{Born1954}
}
\hfill
\parbox{0.23\textwidth}{
$\omega = 2\sqrt{\frac{K}{m}}|\sin(ka/2)|$
}
\hfill
\parbox{0.23\textwidth}{
Periodic in $k$\\
$v_g=0$ at boundaries
}
\hfill
\parbox{0.23\textwidth}{
Acoustic phonons\\
Brillouin zones
}
\vspace{3pt}

\hrule
\vspace{3pt}

\noindent
\parbox{0.23\textwidth}{
\textbf{Diatomic Chain}\\
$m_1 \neq m_2$\\
\cite{Kittel2005}
}
\hfill
\parbox{0.23\textwidth}{
$\omega^2 = \frac{K(m_1+m_2)}{m_1m_2}[1 \pm \sqrt{1 - \frac{4m_1m_2}{(m_1+m_2)^2}\sin^2(ka/2)}]$
}
\hfill
\parbox{0.23\textwidth}{
Two branches\\
Acoustic + optical
}
\hfill
\parbox{0.23\textwidth}{
Band gap\\
IR active modes
}
\vspace{3pt}

\hrule
\vspace{3pt}

\noindent
\parbox{0.23\textwidth}{
\textbf{KdV linearized}\\
$\eta_t + c_0\eta_x + \beta\eta_{xxx} = 0$\\
\cite{Drazin1989}
}
\hfill
\parbox{0.23\textwidth}{
$\omega = c_0k - \beta k^3$
}
\hfill
\parbox{0.23\textwidth}{
Cubic\\
$v_g < c_0$
}
\hfill
\parbox{0.23\textwidth}{
Weak dispersion\\
Soliton foundation
}
\vspace{3pt}

\hrule
\vspace{3pt}

\noindent
\parbox{0.23\textwidth}{
\textbf{Helmholtz}\\
$\nabla^2\psi + k_0^2\psi = 0$\\
\cite{Jackson1998}
}
\hfill
\parbox{0.23\textwidth}{
$k^2 = k_0^2$\\
(fixed $\omega$)
}
\hfill
\parbox{0.23\textwidth}{
Constant $|k|$\\
Eigenvalue
}
\hfill
\parbox{0.23\textwidth}{
Cavity modes\\
Waveguide resonances
}

\end{minipage}
\vspace{2pt}
\hrule

\end{table*}

\section*{Data Availability}

No new data were created or analyzed in this study.

\end{document}